# The Case for Strategic Data Stewardship

Re-imagining Data Governance to Make Responsible Data Re-use Possible

Stefaan Verhulst

**Abstract**

As societal challenges grow more complex, access to data for public-interest use is paradoxically becoming more constrained. This emerging "data winter" is not simply a matter of scarcity, but of shrinking legitimate and trusted pathways for responsible data reuse. Concerns over misuse, regulatory uncertainty, and the competitive race to train AI systems have concentrated data access among a few actors while raising costs and inhibiting collaboration.

Prevailing data governance models—focused on compliance, risk management, and internal control—are necessary but insufficient. They often result in data that is technically available yet practically inaccessible, legally shareable yet institutionally unusable, or socially illegitimate to deploy.

This paper proposes strategic data stewardship as a complementary institutional function designed to systematically, sustainably, and responsibly activate data for public value. Unlike traditional stewardship, which tends to be inward-looking, strategic data stewardship focuses on enabling cross-sector reuse, reducing missed opportunities, and building durable, ecosystem-level collaboration. It outlines core principles, functions, and competencies, and introduces a practical Data Stewardship Canvas to support adoption across contexts such as data collaboratives, data spaces, and data commons.

Strategic data stewardship, the paper argues, is essential in the age of AI: it translates governance principles into practice, builds trust across data ecosystems, and ensures that data are not only governed, but meaningfully mobilized to serve society.

**Introduction**

We are entering a paradoxical moment in the evolution of data-driven societies. At the exact moment that complex societal challenges (e.g. climate adaptation, public health, democratic resilience, societal inequities) demand more robust evidence and analytical capacity, access to data for public-interest use is increasingly constrained and fragmented. This is not because data are scarce, but because legitimate, trusted, and socially sanctioned pathways for using data beyond organizational silos are eroding. These barriers are emerging, in part, as a defensive response to the growing [weaponization of data and openness](#) itself in the race to develop and deploy AI systems.

This contradiction marks the onset of what I have described as a [data winter](#): a period characterized not by the absence of data but by shrinking access for science, policy, and public problem-solving; rising transaction and compliance costs; and growing reluctance among data holders to facilitate reuse beyond tightly controlled environments. At the same time, data access is becoming increasingly concentrated in the hands of a small number of powerful actors, exacerbating asymmetries between those who generate data, those who control it, and those tasked with acting in the public interest.

In practice, this means that humanitarian actors are unable to access mobility data during times of crisis, public health agencies are locked out of social data needed to detect emerging pandemics, and civil society organizations are priced out of data partnerships. The data exist—but the pathways to re-use them are increasingly frozen.

Against this backdrop, traditional approaches to data governance, while necessary, are no longer sufficient. Many remain oriented toward internal control, risk minimization, protection and compliance rather than toward enabling responsible data re-use for societal benefit. As a result, data may be formally governed yet practically inaccessible; legally shareable yet institutionally unusable; technically available yet socially illegitimate to deploy.

It is in this context that the need for what I call *strategic data stewardship* emerges: not as a call for more simplification or deregulation, but as an institutional response to the growing gap between data's potential public value and society's capacity to realize it responsibly.

Strategic data stewardship builds on existing notions of [data stewardship](#) but extends them outward from the organization to the ecosystem. It defines a function and role focused not merely on controlling data, but on enabling responsible, sustainable, and purpose-driven data reuse in the public interest.

In what follows, I examine the limitations of current approaches to data governance and provide an overview of existing typologies and methods of data stewardship. I then build on these notions to describe strategic data stewardship as a systematic approach to enabling responsible data reuse. In addition, I provide a practical "data stewardship canvas" to help organizations translate the concept into operational capabilities, roles, and processes.

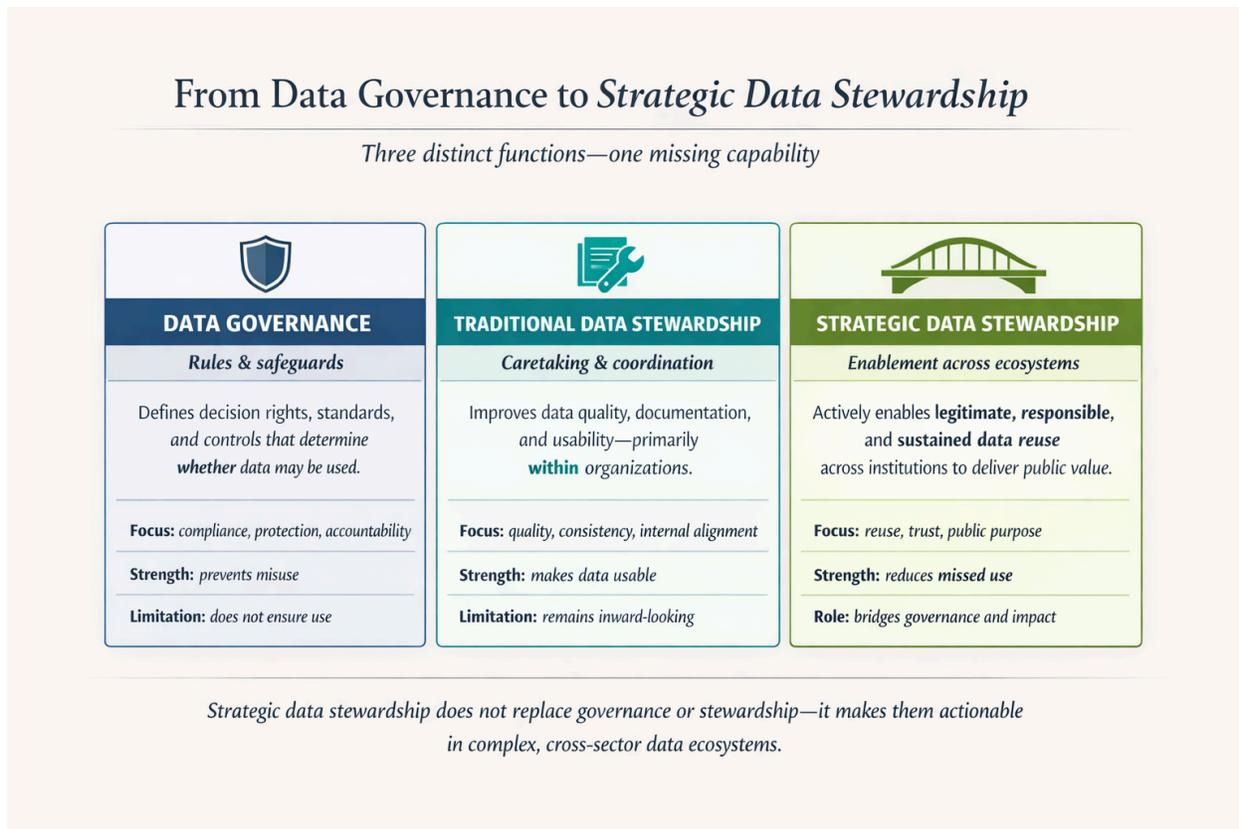

The specific contribution of this paper—articulating and operationalizing strategic data stewardship as a distinct institutional function—is also part of a broader goal to [re-imagine data governance for the age of AI.](#) If AI increases the potential value of data reuse, then a data winter simultaneously increases the risk around that reuse. Bridging this gap requires moving beyond simply more or "better" rules; it requires new competencies and capacities to build trust and manage risk across complex data ecosystems. This, we argue, is the value that can be offered by strategic data stewardship.

I. **Revisiting Data Governance**

The notion of data governance emerged in the late 1990s and early 2000s, alongside the rise of enterprise data management, data warehousing, and regulatory and protection regimes concerned with privacy and security. Done right, [data governance](#) provides the strategic oversight and decision-making framework that aligns processes, practices and technologies around shared principles to meet a particular purpose (the [4Ps of data governance](#))—such as ensuring that data use respects rights, builds trust and delivers measurable social and economic value. It provides the rules, decision rights, standards, and oversight mechanisms that determine how data are collected, processed, shared, analyzed and used (data lifecycle).

Yet despite their undeniable value and even necessity, governance frameworks remain often, by design, conservative. In most cases, their primary function is to reduce risk, ensure compliance, and maintain institutional control. As a result, governance often becomes inward-looking: focused on internal quality, ownership, classification, and access controls. While this reduces potential harm, it also limits potential and prioritizes avoiding [misuse over *missed* use](). It does little to answer a more fundamental question: How do we actually make data more accessible and reusable--at scale and across sectors--for solving societal problems?

***"The central failure of contemporary data governance is not that it protects too much, but that it enables too little."***

This gap between protection and use has widened as data ecosystems have grown more complex. The central failure of contemporary data governance is not that it protects too much, but that it enables too little. In practice, this means that organizations often default to non-sharing, not because data reuse is prohibited, but because no function is explicitly responsible for making it feasible, legitimate, and worthwhile. As data-intensive collaborations increasingly span sectors and jurisdictions, this absence becomes a structural impediment. Increasingly, it is clear that we need a more positive--enabling yet responsible--approach to how institutions manage and mobilize their ever-growing stocks of data.

## II.    Data Stewardship: Potential and Limitations

In recent years, data stewardship has emerged as a potential alternative—or complement—to traditional data governance, particularly in response to the latter's increasingly evident limits. As data ecosystems have grown more distributed and cross-sectoral, a range of stewardship practices have emerged to bridge gaps between data holders and data users, and across institutional boundaries. In [*Data Stewardship Decoded*](), I described this evolution as the emergence of distinct yet complementary manifestations of data stewardship: as a set of competencies and skills, as an organizational role, as an intermediary function, and as a set of guiding principles. [Data collaboratives]() that convene public and private actors around shared problems exemplify stewardship as an intermediary organization while stewardship roles embedded within institutions reflect a growing recognition that data governance requires dedicated functions oriented toward reuse and public value. Across these contexts, data stewardship represents a shift away from purely defensive, compliance-driven data practices toward more intentional, value-oriented approaches that actively enable responsible data reuse, collaboration, and impact.

Similarly, in "[Wanted: Data Stewards]()," my collaborators and I proposed a redefinition of data stewardship that moves beyond narrow technical or business connotations toward a public-value-oriented leadership function. In this conception, data stewardship is not merely about caretaking data assets, but about systematically, sustainably, and responsibly enabling their reuse for societal benefit. Rather than treating data as an internal resource to be guarded, stewardship emphasizes the conditions under which data can be shared, combined, and mobilized to address real-world problems—all while maintaining trust, legitimacy, and accountability.

This redefinition rests on three core responsibilities that cut across sectors and domains (as also recognized by the Canadian Digital Governance Standards Institute Standard [CAN/CIOSC 100-7 | Digital Governance – Part 7, Operating Model for Responsible Data Stewardship](#)):

1. **Collaborate**: Engaging with external partners to identify meaningful questions and opportunities where data reuse can create value.
2. **Protect**: Ensuring that data reuse does not cause harm, respects rights, and aligns with ethical, legal, and societal expectations.
3. **Act**: Closing the loop from data to insight to impact by ensuring that insights are translated into decisions, policies, or services.

What distinguishes this conception of a reimagined stewardship from traditional governance roles is its outward-looking, action-oriented posture. The data steward is not simply a guardian of quality or compliance, but a catalyst within the data ecosystem.

Yet even data stewardship, at least as it is currently conceived and institutionalized, has important limitations. Existing stewardship roles tend to fall into three categories (see figure below):

- Business or domain stewards, focused on data quality and internal use
- Technical stewards or custodians, responsible for infrastructure and security
- Enterprise or coordinating stewards, aligning standards across units

What they share is an inward-facing mandate. What they lack is responsibility for enabling reuse beyond the organization.

| Stewardship Roles | Description | Example Functions |
|---|---|---|
| **Business Data Steward** (other names include Operational Data Steward, Subject Area Steward) | A subject-matter expert (SME) with designated responsibilities for classification, protection, quality, and effective use of one or more data sets | • Define data needs within their functional areas<br>• Understand and answer questions about the meaning, derivation, quality requirements, and uses of data within their scope<br>• Define business rules and data quality objectives<br>• Work to enhance the quality and value of data to the organization<br>• Document data within their scope |
| **Data Custodian or Technical Data Steward** | An individual with physical custody of the data, typically an information technology professional with skills needed to perform data processing tasks | • Assist with technical data documentation<br>• Perform data manipulation including loading, transfer, transformation, and integration<br>• Implement data confidentiality and security requirements including access controls<br>• Perform or monitor data backup and recovery processes<br>• Create and run data profiling and validation scripts; assist with data quality defect resolution |
| **Coordinating Data Steward** (other names include Data Domain Steward and Enterprise Data Steward) | Designated point person for data within a particular organizational unit, function, data domain, or subject area. Responsible for facilitating data improvements and issue resolution across business units; often serves as a liaison to the data governance body | • Identify and communicate data needs, opportunities, and issues for their domain to be addressed through data governance<br>• Coordinate implementation of data standardization and improvement efforts within a major organizational unit (e.g., division or branch)<br>• Identify and support business data or subject area stewards within their designated area<br>• Facilitate cross-divisional collaboration on data improvements<br>• Draft and review proposed new policies, standards, processes, and guidelines<br>• Resolve cross-cutting data issues that go beyond an individual business unit or system<br>• Support data governance compliance<br>• Advocate for data quality and documentation initiatives |

Figure 1. Taxonomy of Data Steward Roles and Function (National Academies)

Even the most advanced "enterprise" or "coordinating" data steward remains largely focused on internal alignment: harmonizing standards, resolving disputes across business units, and supporting governance compliance. As a result, stewardship often inherits the same inward-looking orientation that constrains governance, despite its more expansive aspirations.

What is largely missing is a role explicitly mandated to *accelerate strategic access to data for reuse beyond organizational boundaries* while maintaining legitimacy, trust, and responsibility over time. This gap becomes particularly acute in contexts such as data collaboratives, data spaces, and data commons, where no single actor controls the system, but many depend on its effective functioning. It is this gap between stewardship's promise and its current institutional expression that motivates the need for what I describe as strategic data stewardship.

In this context, the term *strategic* is intentional. It signals a shift from stewardship as a supporting or technical function to stewardship as a core institutional capability, one that proactively shapes priorities, incentives, and long-term value creation, rather than reacting to requests on an ad-hoc basis.

III. **The Strategic Data Steward: Principles and Functions**

Strategic data stewardship emerges as a response to the growing gap between data governance and meaningful data access and reuse, particularly in an environment where data are valuable and contested. It seeks to move beyond ad hoc data sharing arrangements and to make responsible data reuse more systematic, durable, and legitimate across institutional boundaries and sectors. Building on both established data governance and traditional data stewardship practices, strategic data stewardship introduces an explicitly *outward-facing, purpose-driven function focused on enabling reuse at scale rather than merely managing data within organizations.*

At a high level, the strategic data steward can be understood as a functional counterpart to the enterprise data steward, but with a fundamentally different objective. Rather than just optimizing for internal consistency, the strategic data steward focuses on unlocking trusted and responsible data reuse across organizations and sectors. Three core principles guide strategic data stewardship:

- **Systematic**: embedded in repeatable processes rather than ad-hoc exceptions;
- **Sustainable**: supported by viable incentives or business models, resources, and long-term governance arrangements, and environmental impact assessments; and
- **Responsible**: grounded in ethical reflection, risk mitigation, and social license.

What does all this mean in practice? Organizations managing large datasets may be less concerned with formal titles than with functions: what tasks need to be performed, by whom, and how those tasks can be coordinated to enable safe, effective, and legitimate data reuse. Strategic data stewardship therefore shifts attention from individual roles to institutional capacity.

> **A note on terms: governance, stewardship, and what "strategic" adds**
>
> To be clear, strategic data stewardship is not a rebranding of existing data governance or compliance functions, nor is it simply a more senior version of traditional data stewardship. Each plays a distinct role—yet they operate at different levels of intent and action.
>
> **Data governance** establishes the *rules of the game*. It defines decision rights, standards, safeguards, and accountability mechanisms to ensure that data are handled lawfully, securely, and consistently. Governance answers the question: *Under what conditions may data be used at all?* Its strength lies in protection and control; its limitation is that, on its own, it rarely ensures that data are actually reused for meaningful public outcomes.
>
> **Traditional data stewardship**, by contrast, focuses on *caretaking and coordination*. Stewards improve data quality, documentation, and interoperability; they resolve issues across teams and domains; and they help data move within organizations more effectively. Stewardship answers the question: *How can data be managed responsibly and made usable?* Yet most stewardship roles remain inward-looking, embedded within organizational boundaries rather than oriented toward cross-sectoral reuse.
>
> **Strategic data stewardship** adds a missing layer: enablement across ecosystems. It is explicitly concerned with making data reuse *possible, legitimate,* and *durable* beyond organizational silos. Strategic data stewardship answers a different question altogether: *How do we intentionally mobilize data—across institutions and over time—to address societal challenges while maintaining trust, legitimacy, and accountability?* It shifts the focus from controlling data to *activating it responsibly*, from avoiding misuse to reducing missed use, and from ad-hoc sharing to systematic public-value creation.

To move from theory to practice, we developed a Data Stewardship Canvas, increasingly used in setting up data collaboratives, to translate stewardship into concrete institutional capacity. The Canvas seeks to help operationalize strategic data stewardship by focusing on a set of interlocking functions rather than a single job description.

# The Data Stewardship Canvas

Designed by: | Date: | Version:

The Data Stewardship Canvas is a step by step process that maps a data steward's journey when building a data collaborative to support data re-use—whether the data steward is requesting or providing access to data. The steps of the canvas seek to create a systematic and responsible approach to effectively re-using data for positive social and economic outcomes.

## 1. Defining the Demand for Data

- **Framing the problem**: What societal, economic, or organizational challenge are you addressing?
- **Decision mapping**: What decisions do you seek to inform and when in the decision lifecycle is the data needed?
- **Identifying stakeholders**: Who needs to make decisions? Who will act on the data?
- **Question formulation**: What (type of) question, if answered, will inform the decision or address the problem?
- **Bringing all together**: What is your theory of change?

## 2. Defining the Supply of Data

- **Determining the Minimal (Viable) Data Needed**: What are the data elements that are required to answer the priority questions?
- **Data scouting and cataloguing**: What data sources or products exist that match the requirements?
- **Data wrangling and preparation**: How to prepare data to make it reusable?
- **Data audit and tagging**: Assessing and categorizing data in terms of quality, timeliness, interoperability, and ethical or legal considerations.
- **Data ops**: What are the expertise and capacity needs for this project?
- **Addressing data barriers**: Is there a role for synthetic data, proxies, and modeling?

## 5. Matching Demand & Supply: Operational Models

- **Assessing the level of Conditionality for Data Access**: What level of openness or control is appropriate for this project? What mechanisms (e.g. tiered access, data licensing, APIs) will enable controlled access?
- **Designing a Fit-for-Purpose Collaborative Model**: What is the best collaboration model for this data initiative?

## 6. Matching Demand & Supply: Governance

- **Defining the 4 Ps**: How do Purpose, Principles, Processes, and Practices guide governance?
- **Decision Provenance**: Who is responsible and accountable across the data life cycle?
- **Establishing a Social License**: How do we build public trust and legitimacy in data re-use? What engagement methods e.g., consultations, co-design workshops) will foster inclusivity?
- **Compliance Assessment**: What key legal frameworks apply?
- **Operationalization**: How do we translate governance principles into real-world decision-making and enforcement, such as data sharing agreements?

## 7. Matching Demand & Supply: Tech Infrastructure

- **Infrastructure Requirements**: What are the options and requirements regarding Data Transfer, Storage, and Access?
- **Preparing AI-ready data**: What data formats, labeling techniques, and governance are essential for AI applications?

## 4. Assessing the Risk

- **Assessing Risks Across the Data Lifecycle**: What are the risks at each stage of the data lifecycle?
- **Due Diligence of Possible Partners**: What are the risks of providing or receiving access to data for particular stakeholders?
- **Assessing the Risks of Not Having Access to Data**: What critical decisions cannot be made without this data?
- **Externalities Assessment**: What are the intended and unintended consequences of this data initiative?

## 8. Moving from Insight to Action: Decision Intelligence

- **Decision Intelligence**: How do we transform data insights into actionable policy decisions?
- **Lived Experience**: How do we design effective feedback loops?
- **Decision Legitimacy**: How do we embed trust in translation (such as simulations and visualizations) and decision intelligence systems?

## 3. Making a Value Proposition

- **Defining the Value to Society**: What societal problem does this data project address?
- **Identify Beneficiaries**: Who are the primary beneficiaries, and how will they benefit?
- **Making the Business Case for Data Holders**: What incentives do data holders have to participate?
- **Developing a Cost/Benefit estimate**: What are the direct and indirect costs of implementing this project? What are the expected short-term and long-term benefits?

## 9. Measuring Impact

- **KPIs, Impact and Evaluation**: How will you capture the impact and success of this project?
- **Exit Strategies**: How will you know when to end this project? What indicators signal completion, scale-up, or pivoting? How do we define data project sustainability?

Figure 2: The Data Stewards Canvas (Stefaan Verhulst, 2025)

While the specific labels may vary across organizations, the core functions and competencies include:

- **Purpose and Problem Definition**: Identifying priority questions and societal needs that justify data reuse.

- **Data Assessment and Readiness**: Understanding what data exist, their limitations, and what transformations are required for safe reuse.
- **Stakeholder and Partner Engagement**: Convening and brokering relationships among data holders, users, and affected communities.
- **Risk, Ethics, and Protection**: Anticipating harms, embedding safeguards, and aligning with legal and ethical norms.
- **Access Design and Governance**: Designing appropriate access modalities—from secure analysis environments to federated or mediated access.
- **Value Realization and Action**: Ensuring insights inform decisions and that benefits are distributed fairly.
- **Learning and Adaptation**: Continuously evaluating outcomes and impact, legitimacy, and relevance over time.

Seen through the lens of this canvas, strategic data stewardship is not a single task but a portfolio of strategic capabilities that complement formal data governance and data management structures. While governance establishes access rules and accountability, strategic data stewardship focuses on making those rules workable in practice. Together, they form complementary layers of a mature data ecosystem, one that is not only well governed, but capable of delivering sustained public value in the age of AI.

## IV. Conclusion

Conventional wisdom today holds that we are still at a very early stage of AI innovation, a formative period in which core architectures, norms, and institutional arrangements are only beginning to take shape. If this is true, then the role of data—and by extension data stewardship—is likely to become increasingly important in the years and decades ahead. This is particularly the case given several underlying structural forces both driving and being accelerated by the advent of AI. These include the growing concentration of high-value data and compute resources; rising concerns about extraction, privacy, bias, and misuse; increasing geopolitical and commercial competition over data assets; and expanding demand for data reuse to address complex societal challenges. Each of these forces requires careful calibration between access and protection, innovation and accountability, and private incentives and public value.

Existing data governance structures are not well suited to this task. Crafted a quarter-century ago when data were smaller in scale and primarily used within organizations, they are ill-equipped for an era of large-scale data collaboration and cross-sectoral reuse. If we do not complement governance frameworks with data stewardship, we risk entrenching fragmentation, reinforcing data silos, and deepening digital and social asymmetries. This would represent a missed opportunity not only for AI innovation, but for the ability of societies to use data in service of shared goals, democratic accountability, and public problem-solving.

Strategic data stewardship offers a way forward. By embedding stewardship as a recognized institutional function--alongside governance, IT, and security--organizations can move from reactive, risk-averse data sharing toward purposeful, legitimate, and sustained data reuse. Strategic data stewards help translate governance principles into practice, align data use with clearly articulated societal objectives, and build the social license and coordination required for data ecosystems to function over time. In doing so, they provide a practical pathway to ensure that data are not only well governed, but meaningfully mobilized to deliver lasting public value in the age of AI.

The question is no longer whether societies need more data. They do. The question is whether we will build the institutional capacity to re-use that data responsibly, at scale, and in service of public value. Strategic data stewardship is one such capacity; and one we can no longer afford to treat as optional.

**Further Reading:**

Verhulst, S. (2025). [Data stewardship decoded: Mapping its diverse manifestations and emerging relevance at a time of AI](#) (SSRN Scholarly Paper No. 5124555). Social Science Research Network.

Verhulst, S., & Benjamins, R. (2024, June 13). [Misuse versus missed use — The urgent need for chief data stewards in the age of AI. Medium](#). Data & Policy Blog.

Verhulst, S. G. (2023, March 13). [Wanted: Data stewards — Drafting the job specs for a re-imagined data stewardship role.](#) Medium. Data Stewards Network.

Verhulst, S. et al (2020). [Wanted: Data stewards — (Re-)Defining the roles and responsibilities of data stewards for an age of data collaboration (Position Paper)](#)

**About the author:**

Dr. Stefaan Verhulst is Co-Founder of The GovLab (New York City) and The DataTank (Brussels). He is also Research Professor at the Tandon School of Engineering, New York University, and Editor-in-Chief of the journal *Data & Policy* (Cambridge University Press). His work focuses on designing governance frameworks and institutional capacities that enable the responsible use of data and AI in the public interest.